\documentclass[preprint,12pt]{elsarticle}
\usepackage{amsthm}
\usepackage{graphicx,natbib,amssymb}
\usepackage{newlfont}
\usepackage{graphicx}
\usepackage{subfigure}
\usepackage[section]{placeins}
\usepackage{psfrag}
\usepackage{color}
\usepackage{caption2}
\usepackage{subeqn}
\usepackage{flafter}
\usepackage{bm}%
\def\fnum@figure{\figurename\thefigure}
\renewcommand{\figurename}{Fig.}

\journal{Physics Letters A}
\begin{document}
\title{Self-trapped dynamics of a hollow Gaussian beam in metamaterials}
\author{A. K. Shafeeque Ali and M. Lakshmanan$^\ast$}
 \address{Department of Nonlinear Dynamics, School of Physics, Bharathidasan University, Tiruchirappalli-620 024, India}
 \begin{abstract}
We present a systematic investigation on the dynamics of a hollow Gaussian beam (HGB) in metamaterials. We predict self-trapped propagation of HGBs and evolution of the beam is highly influenced by dimensionless dispersion coefficient ($\kappa$), which determines the strength of dispersion over diffraction.  The evolutions of HGBs such as disappearance of single ringed intensity pattern and appearance of patterns with a central bright spot are achievable with less propagation distance in  metamaterials with higher values of $\kappa$. On the other hand, metamaterials with low values of $\kappa$  can preserve single ring intensity distribution over a long propagation distance without focusing. When the strength of dispersion over diffraction increases, it significantly influences the evolution of the beam and may lead to the formation of tightly focussed beam with high peak intensity at the center. The phenomenon of tight focussing is found to have some applications in trapping of nanosized particles.
\end{abstract}
\maketitle
\section{Introduction}
Recently, hollow gaussian beams (HGBs) have aroused increasing research interest and extensive attention has been given due to their potential applications in the fields of binary optics, atomic optics, optical trapping and optical communication \cite{tak,Jia}. A mathematical model to describe dark hollow beam (DHB) has been proposed and its propagation properties in free space have been analyzed \cite{sinh}. Also the dynamics of HGBs has been studied in many systems, namely in uniaxial crystals \cite{Uni}, plasma \cite{Ms}, thin metal films \cite{Ms1},  strongly nonlocal nonlinear media \cite{nonlocal,nonlocal1} and so on.
 \begin{figure*}
\begin{center}
    \subfigure[]{\label{abc1}\includegraphics[height=4.5 cm, width=5 cm]{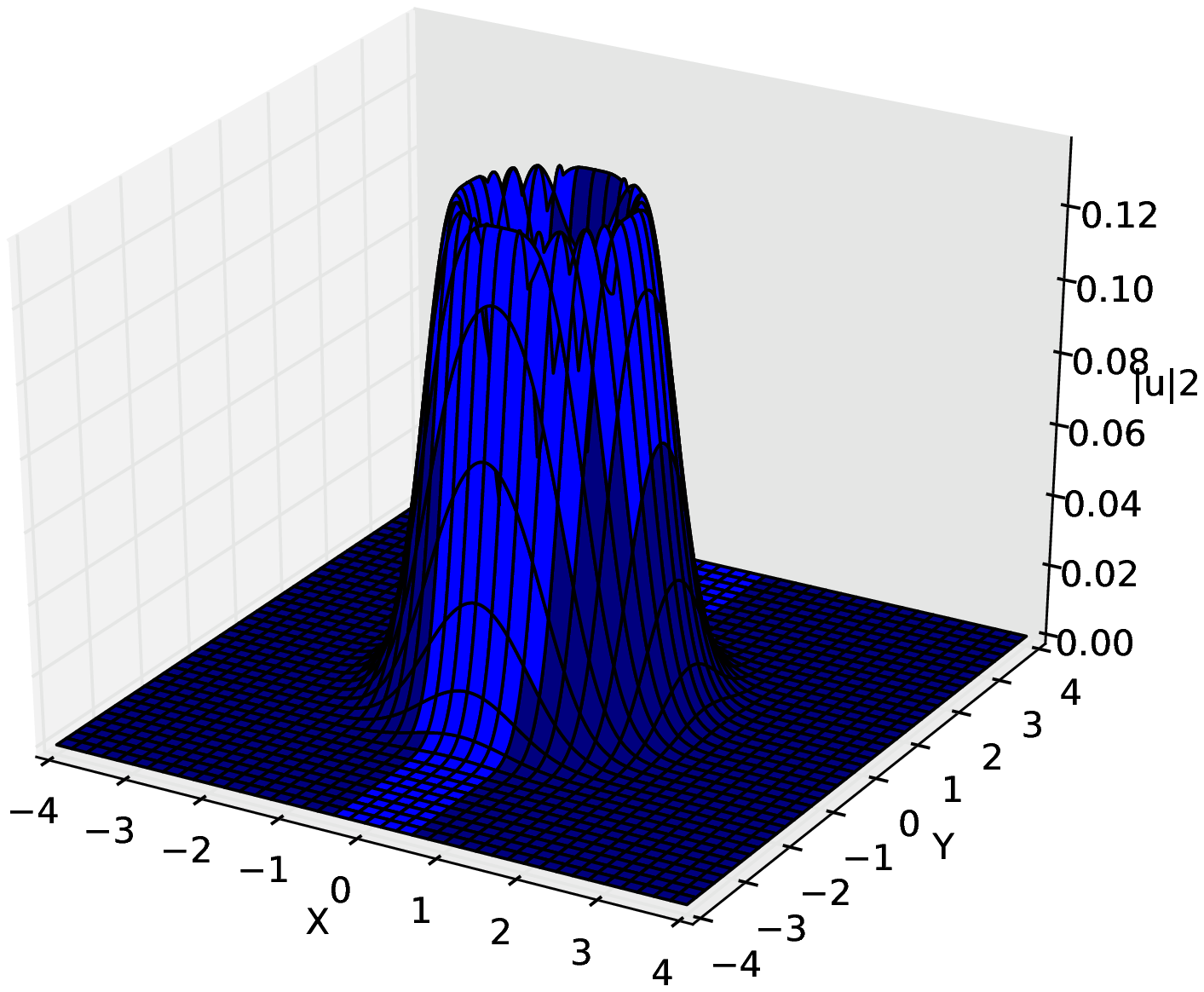}}
    \subfigure[]{\label{abc1}\includegraphics[height=3.5 cm, width=3.5 cm]{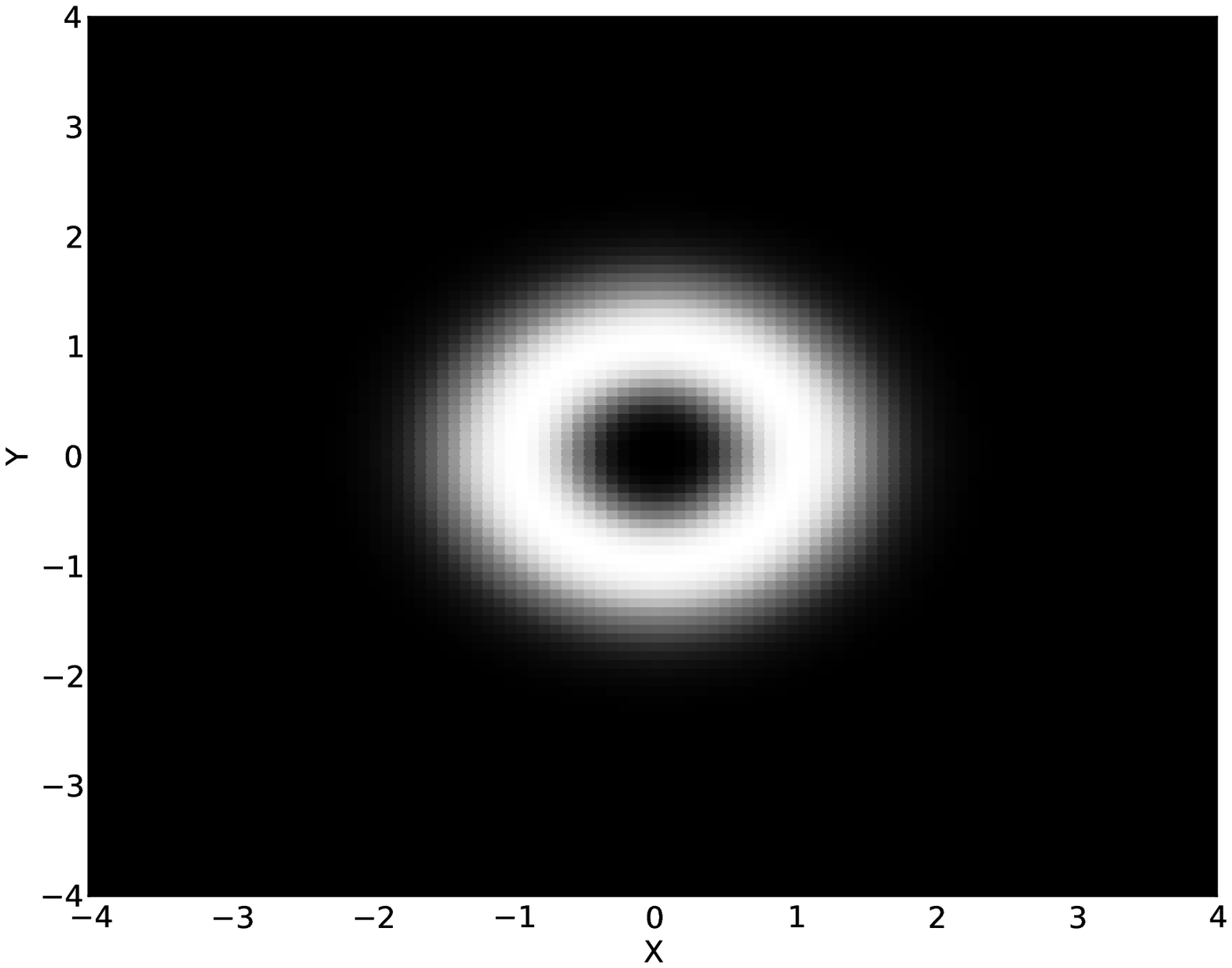}}
\caption{(Color online) First order hollow Gaussian beam.}
  \label{beamm}
 \end{center}
\end{figure*}
HGBs can be generated by spatial filtering \cite{G1}, phase-only filtering \cite{G2}, nonlinear interaction of photons with orbital angular momentum  \cite{G3}, beam shaping with phase-only liquid crystal spatial light modulator \cite{G4} and so on. The second harmonic generation as a result of self focussing HGBs  in collisionless plasma has been analyzed  \cite{SH}.  The radiation force on a dielectric  sphere  produced by HGBs  in  the  Rayleigh scattering regime has been investigated and the possibility to guide the particles with refractive index lower than the ambient has been demonstrated \cite{trap}. The  highly focused HGBs can be used to trap particles with low or high index of refraction \cite{trap1}. The tight focused HGBs can be used to trap and manipulate nanosized dielectric spheres with high-refractive index in the focal region \cite{trap2}.
\par
On the other hand, the manipulation of light propagation properties with metamaterials has opened new interesting fields of research \cite{mt1, mt2, mt3}. This new class of artificial materials show uncommon and exotic electromagnetic characteristics such as negative refraction, reversed Doppler shift, reversed Goos-Hanchen shift, reversed Cerenkov radiation and reversal of Fermat's principle, super-resolution imaging and invisibility cloaking \cite{Ramakrishna, Caloz}. The metamaterials not only operate in an invisible range of frequencies but also their application in the optical frequency range has been developed \cite{opt}. {\color {blue}Designing of NIM becomes practical with the exciting idea of Pendry \emph{et al.} \cite{Pendry1, Pendry2, Pendry3} who offered the way towards the complete controllability of electric permittivity and magnetic permeability by using the geometrical parameters of composite elements. As a consequence, the thin wire lattice and split ring resonator are the two basic building blocks of NIM medium where the former gives the negative electric response and the latter gives the negative magnetic response. Following this remarkable suggestion, Smith \emph{et al.} have achieved the first experimental realization of NIM in the microwave regime, by constructing negative electric permittivity and magnetic permeability components \cite{Smith}. This engineered material with designed inclusion can also exhibit nonlinear effects including harmonic generations \cite{roppo}. The nonlinear NIMs are created by introducing nonlinear elements such as diodes \cite{Ilya} to resonant meta-atoms or by embedding array of meta-atoms into a nonlinear dielectric medium \cite{Zharov}.}
\par
 Investigation on the propagation of electromagnetic waves in the nonlinear regime of metamaterials is being actively developed \cite{n1, n2, n3}. For example second-harmonic generation \cite{n11}, angular multistability and a pronounced directional hysteresis behavior \cite{n12}, controllable Raman soliton self-frequency shift \cite{n13}, bistability  and self-tunability \cite{n14} and modulation instability \cite{n15} have been investigated to mention a few.
\par
It is well known that metamaterials are engineered materials and their electromagnetic properties can be tuned at will. This characteristics make the metmaterials as suitable candidates for many applications. The phenomenon of dispersion and diffraction of the propagating beam through the metamaterials can be tailored properly in such a way that it can significantly change the propagation properties of the beam. In this paper, we theoretically investigate the propagation dynamics of first order HGBs whose profile is shown in Fig. \ref{beamm} in the nonlinear metamaterials. When the strength of the dispersion over diffraction increases, it significantly influences the evolution of the beam and may lead to the formation of tightly focussed beam with high peak intensity at the center. The phenomenon of tight focussing is found to have some applications in the trapping and guiding of the particle.
\par
The rest of the paper is organized as follows. In section II the theoretical model of the problem is presented and the evolution equations for the beam widths are obtained by adopting the Lagrangian variational method. Investigations on the evolution of the HGB in the metamaterial, based on variational as well as numerical analysis on the influence of $\kappa$ on critical power and the phenomenon of tight focusing have been carried out in detail in section III, section IV and  section V, respectively. Conclusions are made in section VI.
\section{Theoretical model}
The propagation of electromagnetic waves in the metamaterials can be described by the following normalized nonlinear Schr\"{o}dinger equation \cite{wenf,wen1},
\begin{equation}
\frac{\partial u}{\partial z }=-i\,\kappa\frac{sgn(\beta_2)}{2}\, \frac{\partial^2 u}{\partial t^2}+ i\,\frac{sgn(n)}{2}\nabla^2_\bot u+i \gamma |u|^{2}\,u.
\label{modeleqn11}
\end{equation}
Here $u(x, y, z, t)$ is the normalized complex amplitude of the propagating mode. $sgn()$ stands for the sign of the particular parameter. $\beta_2$ and $n$ represent group velocity dispersion (GVD) and refractive index, respectively. Hence, sgn$(\beta_2)$ = $\pm1$ corresponds to normal and anomalous
GVDs, respectively, and sgn($n$) = $\pm1$ corresponds to
positive and negative indices of refraction, respectively. $\kappa$ is the
dimensionless dispersion coefficient and $\nabla^2_\bot=\frac{\partial^2}{\partial x^2}+\frac{\partial^2}{\partial y^2}$ is the transverse Laplacian operator. $z$ and $t$ stand for the direction of propagation of the beam and the time in co-moving frame of reference, respectively.   Also, $\gamma$ represents the nonlinear coefficient which can be represented in terms of characteristic length and susceptibilities  as follows,
\begin{equation}
\gamma= sgn(\chi_e^{(3)})\frac{L_{DF}}{L_{ENL}}(1+sgn(\frac{\chi_m^{(3)}}{\chi_e^{(3)}})\frac{L_{ENL}}{L_{HNL}}),
\label{gamma}
\end{equation}
{\color {blue}where $\chi_e^{(3)}$ and $\chi_m^{(3)}$ are third order electric and magnetic susceptibilities, respectively.} $L_{DF}$, $L_{ENL}$ and $L_{HNL}$ are diffraction, electric nonlinear and magnetic nonlinear lengths, respectively. $sgn(\chi_e^{(3)})=\pm1$ stands for focussing and defocussing electric nonlinearities, respectively. $sgn(\chi_m^{(3)})=\pm1$ stands for focussing and defocussing magnetic nonlinearities, respectively. {\color {blue}Even though the magnetic nonlinear response is prominent at microwave range \cite{Zharov},  it is not negligible at other frequencies \cite{CJ}.} It is evident from Eq. (\ref{gamma}) that $sgn(\gamma)$ may be positive or negative as a result of the combined impact of electric and magnetic nonlinearities.
\par
The Lagrangian corresponding to the propagation model given in Eq. (\ref{modeleqn11}) can be written as,
\begin{eqnarray}
\label{leg}
L= \frac{i}{2}\,(u\frac{\partial u^*}{\partial z }- u^* \frac{\partial u}{\partial z })-\,\kappa \frac{sgn(\beta_2)}{2}  |\frac{\partial u}{\partial t }|^2 \nonumber \\ +\,\frac{sgn(n)}{2}  |\frac{\partial u}{\partial x }|^2 +\,\frac{sgn(n)}{2}  |\frac{\partial u}{\partial y }|^2-\frac{\gamma}{2}\,|u|^{4},
\end{eqnarray}
Associated with the above Lagrangian let us now consider the propagation of a HGB  of the following form \cite{beam, beam2},
\begin{eqnarray}
\label{wf11}
u(z,x,y,t)=\Phi(z)((\frac{x}{r a(z)})^{2}+(\frac{y}{r b(z)})^{2}+(\frac{t}{r c(z)})^{2})^m \nonumber \\  e^{i\theta(z)} e^{-((\frac{x}{r a(z)})^{2}+(\frac{y}{r b(z)})^{2}+(\frac{t}{r c(z)})^{2})} e^{i(\alpha(z) x^2+\beta(z) y^2+\delta(z)t^2)},
\end{eqnarray}
where $r$ is a constant and $m=1,2,3,...$ is the order of HGB. We consider in the present paper only the first order HGB corresponding to $m=1$ as Lagrangian variational calculation is more tedious with higher order $(m>1)$ HGBs. We hope to peruse the nature of higher order HGBs in future.   $\Phi(z)$ represents the amplitude of the beam. $a(z)$, $b(z)$ and $c(z)$ are
beam width parameters along  the $x$, $y$ and $t$ directions, respectively.
$\alpha(z)$, $\beta(z)$ and $\delta(z)$ are the parameters to account for the
chirps along $x$, $y$ and $t$, respectively. The reduced Lagrangian for
the system can be calculated using the following form,
\begin{equation}
\langle\pounds\rangle=\int_{-\infty}^\infty \int_{-\infty}^\infty \int_{-\infty}^\infty L dx dy dt.
\label{leff}
\end{equation}
Making use of expression (\ref{wf11}) for $u(z,x,y,t)$ in the Lagrangian (\ref{leg}) and substituting it in Eq. (\ref{leff}), and performing the integrations, we obtain the effective Lagrangian
\begin{eqnarray}
\langle\pounds\rangle=\frac{1}{65536}\frac{r}{a(z) b(z) c(z)}\pi^{3/2}\Phi(z)^2\{5632\sqrt{2}sgn(n)b(z)^2 c(z)^2\nonumber \\ +8960 \sqrt{2} r^4 a(z)^4 b(z)^2 c(z)^2(2 sgn(n) \alpha(z)^2+\frac{\partial \alpha(z)}{\partial z})\nonumber \\ +a(z)^2[5632 \sqrt{2}sgn(n) c(z)^2+8960 \sqrt{2} r^4 b(z)^4 c(z)^2 \nonumber \\ (2sgn(n) \beta (z)^2+\frac{\partial \beta}{\partial z})+b(z)^2 (-5632 \sqrt{2} \kappa sgn(\beta_2)\nonumber \\ -8960 \sqrt{2} r^4 c(z)^4 (2 \kappa sgn(\beta_2) \delta(z)^2-\frac{\partial \delta(z)}{\partial z} )\nonumber \\ +15 r^2 c(z)^2(-63 \gamma \Phi(z)^2+1024 \sqrt{2} \frac{\partial \theta(z)}{\partial z}))]\}.
\label{legran}
\end{eqnarray}
Varying the above effective Lagrangian with respect to the variational parameters given in Eq. (\ref{wf11}) we get the following evolution equations,
\begin{subequations}
\begin{eqnarray}
\alpha(z)=\frac{\partial a(z)}{\partial z}(\frac{1}{2sgn(n) a(z)}),
\label{leg1}
\end{eqnarray}
\begin{eqnarray}
\beta(z)=\frac{\partial b(z)}{\partial z}(\frac{1}{2sgn(n) b(z)}),
 \label{leg2}
 \end{eqnarray}
 \begin{eqnarray}
 \delta(z)=-\frac{\partial c(z)}{\partial z}(\frac{1}{2sgn(\beta_2) c(z)}),
 \label{leg3}
 \end{eqnarray}
  \begin{eqnarray}
 \frac{\partial^2 a(z)}{\partial z^2}= \frac{sgn(n)(11264 \sqrt{2} sgn(n)-945 \gamma r^2 a(z)^2 \Phi (z)^2)}{8960 \sqrt{2} r^4 a(z)^3},
 \label{leg4}
  \end{eqnarray}
  \begin{eqnarray}
 \frac{\partial^2 b(z)}{\partial z^2}= \frac{sgn(n)(11264 \sqrt{2} sgn(n)-945 \gamma r^2 b(z)^2 \Phi (z)^2)}{8960 \sqrt{2} r^4 b(z)^3},
 \label{leg5}
  \end{eqnarray}
  \begin{eqnarray}
 \frac{\partial^2 c(z)}{\partial z^2}= \frac{sgn(\beta_2)(11264 \sqrt{2} sgn(\beta_2)+945 \gamma r^2 c(z)^2 \Phi (z)^2)}{8960 \sqrt{2} r^4 c(z)^3},
 \label{leg6}
  \end{eqnarray}
  \begin{eqnarray}
 P= \frac{15 \pi ^{3/2} r^3 a(z) b(z) c(z) \Phi(z)^2}{32 \sqrt{2}}.
 \label{leg8}
 \end{eqnarray}
\end{subequations}
\begin{figure}[!ht]
\begin{center}
\includegraphics*[height=4.3cm, width = 6cm]{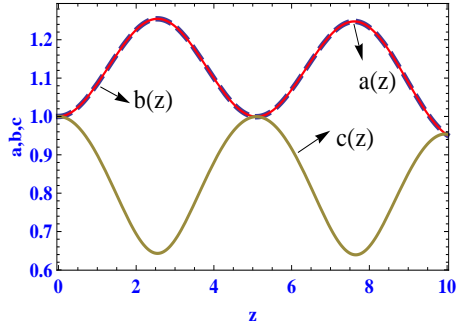}
\caption {(Color online) Periodic variation of beam widths $a(z)$, $b(z)$ and $c(z)$ of HGBs in the negative index regime of metamaterial against the distance of propagation for $\kappa$=0.5. Other parameters are $P_c$ =26.2572, $r=1$, $sgn(\beta_2)=1$ and $\gamma$=-1.}
\label{abc1231a}
\end{center}
\end{figure}
It is interesting to note from Eq. (\ref{leg1}) and Eq. (\ref{leg2}) that the chirps along the x and y coordinates are negative when the HGB propagates in the negative index regime, whereas it is positive in the positive index regime of the metamaterial. The chirp along the temporal coordinate takes a positive or negative value when the dispersion is anomalous or normal, respectively. Also Eq. (\ref{leg8}) shows that $P=\frac{15 \pi ^{3/2} r^3 a(z) b(z) c(z) \Phi(z)^2}{32 \sqrt{2}}$ is a constant of motion and it is conserved during the propagation of the beam. Substituting Eq. (\ref{leg8}) into Eqs. (\ref{leg4}-\ref{leg6}) we obtain the following evolution equations for the beam widths a(z), b(z) and c(z):
\begin{subequations}
\begin{eqnarray}
\frac{\partial^2 a(z)}{\partial z^2}=\frac{44}{35 r^4 a(z)^3}-\frac{9 P \gamma sgn(n)}{40 \pi ^{3/2} r^5 a(z)^2 b(z) c(z)},
 \label{legg1}
 \end{eqnarray}
 \begin{eqnarray}
\frac{\partial^2 b(z)}{\partial z^2}=\frac{44}{35 r^4 b(z)^3}-\frac{9 P \gamma sgn(n)}{40 \pi ^{3/2} r^5 a(z) b(z)^2 c(z)},
 \label{legg2}
 \end{eqnarray}
 \begin{eqnarray}
\frac{\partial^2 c(z)}{\partial z^2}=\frac{44 \kappa^2}{35 r^4 c(z)^3}+\frac{9 P \gamma \kappa sgn(\beta_2)}{40 \pi ^{3/2} r^5 a(z) b(z) c(z)^2}.
 \label{legg3}
 \end{eqnarray}
 \label{legga}
 \end{subequations}
The information about the evolution of the beam width during the propagation of HGB can be obtained from  Eqs. (\ref{legga}). Now we solve above evolution equations numerically to study the dynamics of HGB in metamaterials by imposing some special conditions to uncover self-trapped propagation of the beam. The self-trapped propagation arises with proper choice of parameters in which the solutions to Eq. (\ref{legga}) oscillate periodically.
\section{Evolution of the HGB in the metamaterial}
Here we unfold certain interesting propagation properties of HGB in a nonlinear metamaterial in detail.
The strength of dispersion over diffraction described by the value of $\kappa$ may change the balancing conditions to form soliton-like propagation. The beam evolution as well as self-trapping are also considerably influenced by the value of $\kappa$. Hence in this section, we study the impact of $\kappa$ on the beam evolution in metamaterials. As illustrative examples we consider three different cases ($\kappa$=0.5, $\kappa$=1.0 and $\kappa$=2.0) for our discussion. This theoretical study has physical relevance due to the availability of engineering freedom in metamaterials at will.
\subsection{Case 1: $\kappa$=0.5}
\begin{figure*}
\begin{center}
\includegraphics*[height=12cm, width = 14.5cm]{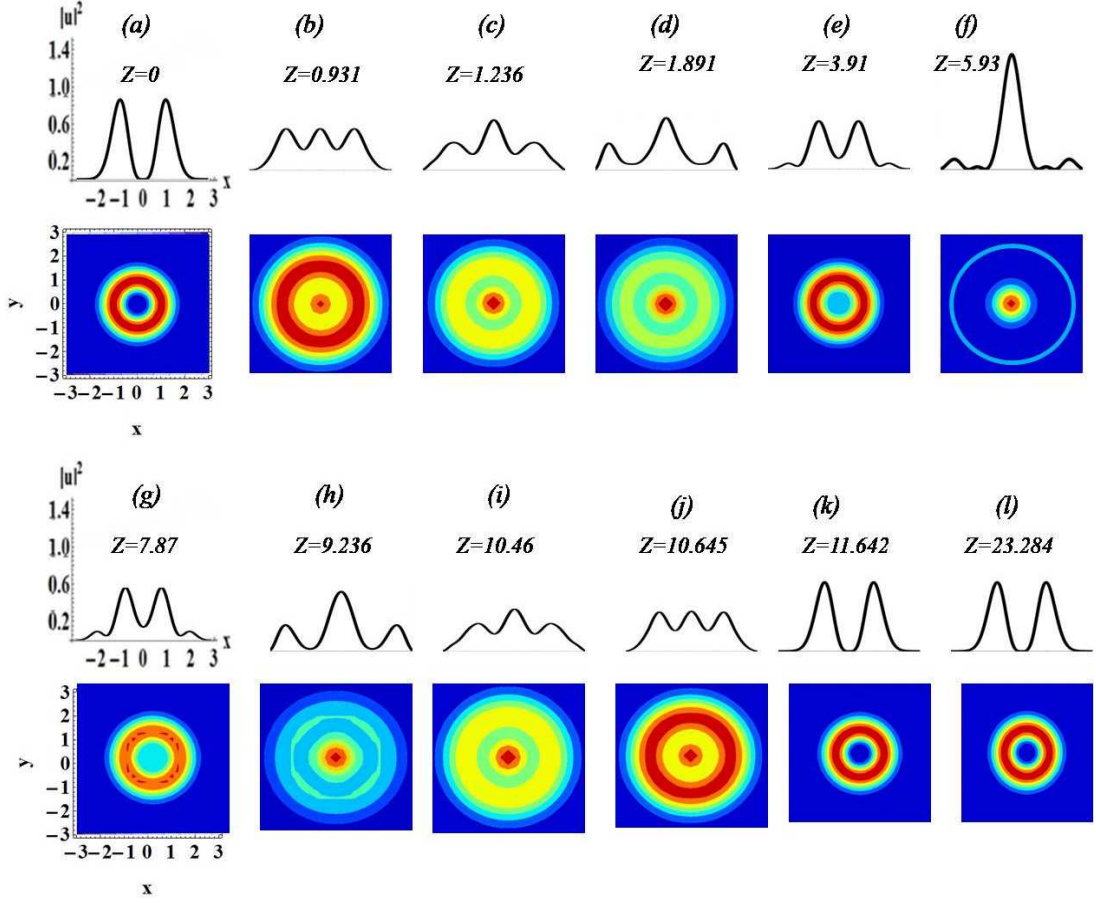}
\caption {(Color online) Evolution of spatial profile of the HGB against propagation distance in the negative index regime of metamaterial for $\kappa$=0.5. First row shows one-dimensional view of the profile and the corresponding  two-dimensional view of the evolution is displayed in the second row. Other parameters are $r=1$, $sgn(\beta_2)=1$ and $\gamma$=-1.}
\label{m1sp5}
\end{center}
\end{figure*}
\begin{figure*}
\begin{center}
\includegraphics*[height=4.3cm, width = 6cm]{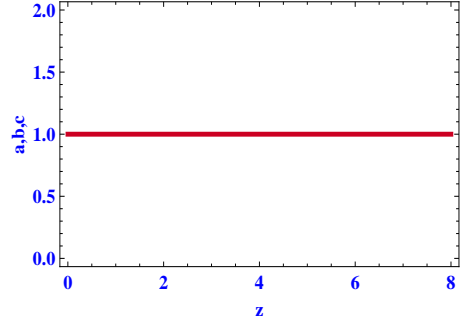}
\caption{(Color online) Beam widths $a(z)$, $b(z)$ and $c(z)$ of HGBs versus the distance of propagation in the negative index regime of metamaterial for $\kappa$=1. Other parameters are $P_c$=32.12, $r=1$, $sgn(\beta_2)=1$ and $\gamma$=-1.}
 \label{m1new}
 \end{center}
\end{figure*}
\begin{figure*}
\begin{center}
\includegraphics*[height=4.3cm, width = 6cm]{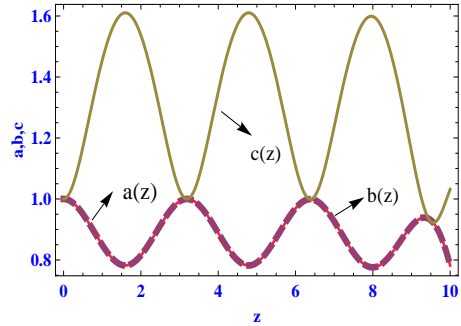}
\caption {(Color online) Periodic variations of beam widths $a(z)$, $b(z)$ and $c(z)$ of HGB in the negative index regime of metamaterial against the distance of propagation for $\kappa$=2. Other parameters are $P_c$ =41.898, $r=1$, $sgn(\beta_2)=1$ and $\gamma$=-1.}
\label{abc1231}
\end{center}
\end{figure*}
\begin{figure*}
\includegraphics*[height=12cm, width = 14.5cm]{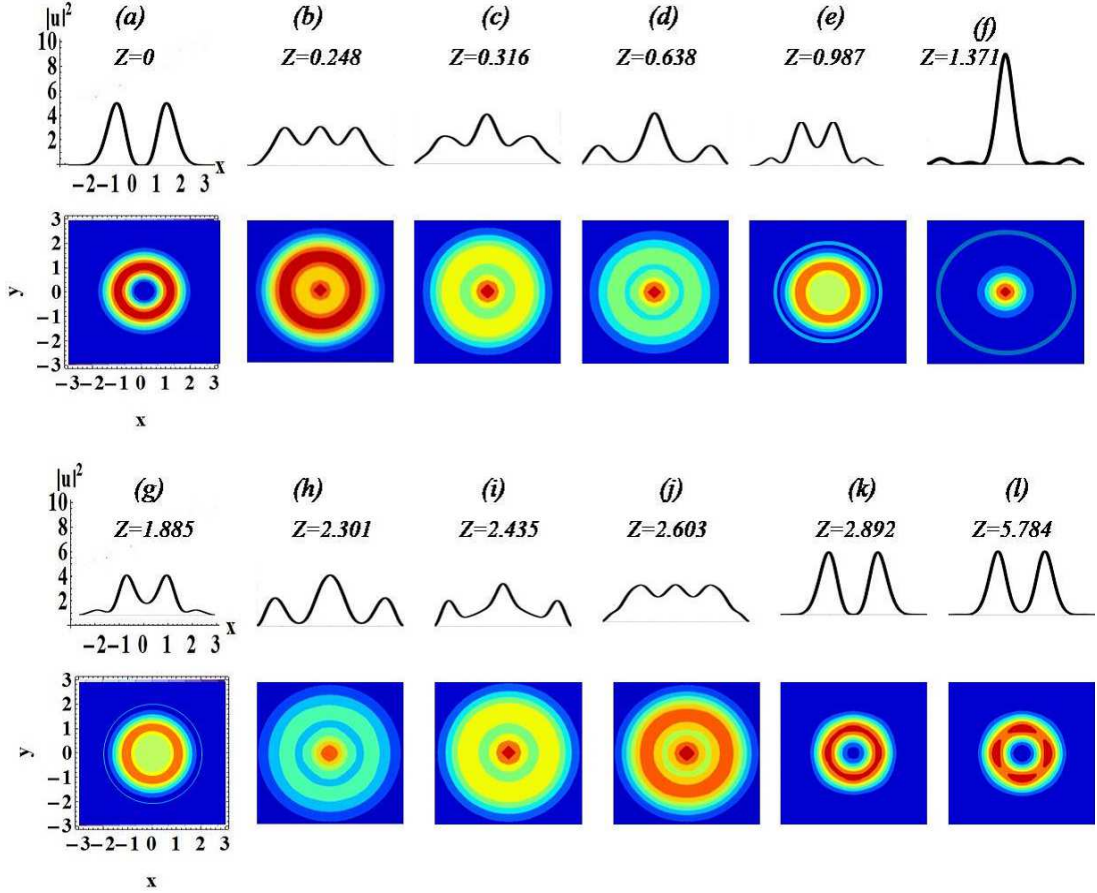}
\caption{(Color online) Evolution of spatial profile of the HGB against propagation distance in the negative index regime of metamaterial for $\kappa$=2. First row shows one-dimensional view of the profile and the corresponding  two-dimensional view of the evolution is displayed in the second row. Other parameters are $r=1$, $sgn(\beta_2)=1$ and $\gamma$=-1. Note the reduction in the period of self-trapping in the present case of $\kappa$=2 compared to the earlier case $\kappa$=0.5 (Fig.\ref{m1sp5}).}
\label{abc1112b}
\end{figure*}
When $\kappa$=0.5, diffraction dominates over the dispersion and there is no delicate balance
among nonlinearity, diffraction, and dispersion to propagate the beam as a soliton-like entity. Even though there is no soliton-like propagation, the beam may get self-trapped at equal intervals of distance of propagation as shown in  Fig. \ref{abc1231a}. The figure depicts how the beam widths vary  to get self-trapped with equal periodicity. This self-trapping phenomenon is observed at a particular input wave power denoted as critical power $(P_c)$. In the present case $P_c$ =26.2572. When $\kappa$=0.5 the effect of dispersion is weaker than the diffraction on the propagating beam. At this condition, during the propagation, the beam width along the temporal coordinate (c(z)) initially decreases, it reaches a minimum, then it increases to the initial value and then it varies periodically. But the beam widths along spatial coordinates (a(z) and b(z)) initially increase, they reach their maximum and then decrease to initial value as shown in Fig. \ref{abc1231a}. This preponderance is attributed to the fact that the action of self focussing nonlinearity on dispersion makes the beam to compress along temporal coordinate initially. Then the width c(z) becomes narrower as the dispersion becomes weaker and weaker and c(z) reaches its minimum point. At this point the dispersion counteracts, becomes stronger and the width c(z) is restored to initial value. On the other hand, the impact of diffraction is stronger than the self focussing nonlinearity and the pulse diverges initially along the spatial coordinates and a(z) and b(z) reach their maximum point. Then the nonlinearity counteracts over diffraction and hence the beam widths a(z) and b(z) get restored to their original values.
\par
It is well known that the self-trapping of light beams in conventional positive refractive index materials may happen for focusing nonlinearity
and anomalous GVD. In the case of metamaterials, however, the circumstance becomes different in the negative index region. It is evident from Eq. (\ref{modeleqn11})  the signs of nonlinearity and dispersion will be reversed as a result of  the negative refractive index. Hence self-trapping of the beam may happen in the normal dispersion regime with defocussing nonlinearity in the negative index regime whereas it is observed in the anomalous dispersion region with focussing nonlinearity in the positive index regime of the metamaterials.
\par
Typical evolutions of HGB in the negative index regime of metamaterial with  defocusing nonlinearity and normal GVD are summarized in Fig. \ref{m1sp5}. The results in these figures and in the following ones are obtained by solving the nonlinear partial differential equation (1) numerically using Crank-Nicolson method. Fig. 3(a) shows the spatial counterpart of the profile at $z=0$. This intensity evolution is also periodic in nature and each period can be divided into two halves (Figs. 3(a)-3(f) and Figs. 3(f)-3(k)).  During first half the beam evolves into a focussed beam and the important stages of evolution are shown in Fig. \ref{m1sp5}.  When the beam propagates the intensity of the beam converges gradually to the center of the beam as depicted in Fig. 3(b). Then the HGB transforms to a beam with high intensity central core and one side lobe (Figs. 3(c) and 3(d)). When it propagates further the central high intensity core splits making low intensity central part and two side lobes in the beam (Fig. 3(e)). After this the central intensity grows again to form a comparatively high focussed beam with two side lobes as depicted in Fig. 3(f). In the second half, the beam evolution becomes reverse to the first and the beam is restored. Fig. 3(l) represents the beam after second period of self-trapping.
\subsection{Case 2: $\kappa$=1.0}
In the case of $\kappa$=1.0 the effects of dispersion and diffraction on the propagating beam will not dominate one over the other and as a result of the adequate balance with nonlinearity the invariance of beam widths supports soliton-like propagation. At a particular critical power the beam widths remain invariant during propagation as depicted in Fig. \ref{m1new}. As a result, the HGB propagates similar to soliton. In this case we have observed $P_c$=32.12. This phenomenon is observed in the normal dispersion regime with defocussing nonlinearity in the negative index regime whereas it occurs in the anomalous dispersion regime with focussing nonlinearity in the positive index cases of the metamaterial.
\par
Although the beam width evolution supports a soliton-like propagation, the intensity evolution is different. The HGB intensity distribution changes continually during the propagation unlike solitons, but it is periodic in the propagation distance similar to Fig. \ref{m1sp5}.
\begin{figure*}
\begin{center}
\includegraphics*[height=4.5cm, width = 6cm]{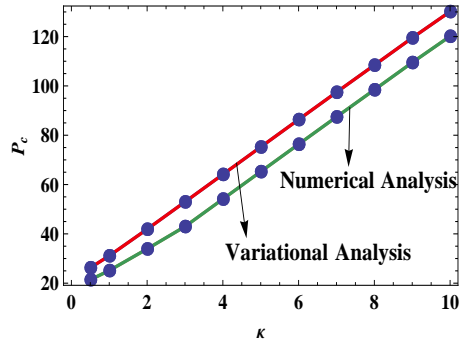}
\caption {(Color online) Variation of critical power ($P_c$) with $\kappa$.}
\label{pcr}
\end{center}
\end{figure*}
\subsection{Case 3: $\kappa$=2.0}
In this case the influence of dispersion on the propagating beam is stronger than the diffraction effect and hence the pulse transforms periodically and gets self-trapped at equal distances as in the earlier case of  $\kappa$=0.5. The distinct periodic oscillations of the beam widths a(z), b(z) and c(z) are depicted in Fig. \ref{abc1231}. In contrast to the case of $\kappa$=0.5, in this instance the value of c(z) grows initially to a maximum point and then it decays to the initial value; however, the reverse is true in the case of spatial beam widths a(z) and b(z). This situation arises due to the fact that the impact of self focussing nonlinearity on strong dispersion makes the beam to diverge in temporal coordinate while the action of nonlinearity on weak diffraction is to make the beam widths a(z) and b(z) to decay initially. The counter action at extremum point results in a reversal of action and brings it back to the initial beam widths. This leads to periodic self-trapping of the HGB in metamaterials.
\par
The typical intensity evolution for the case of $\kappa$=2.0 is depicted in  Fig. \ref{abc1112b}. The evolution stages resemble the case of   $\kappa$=0.5, but the period of self-trapping and profile intensity vary. This variation is  the result of higher self-trapping critical power. As the value of $\kappa$ increases the self-trapping critical power increases, which results in a reduction of the self-trapping period.
\par
\begin{figure*}
\begin{center}
\subfigure[]{\label{abc1}\includegraphics[height=3.7 cm, width=4.5 cm]{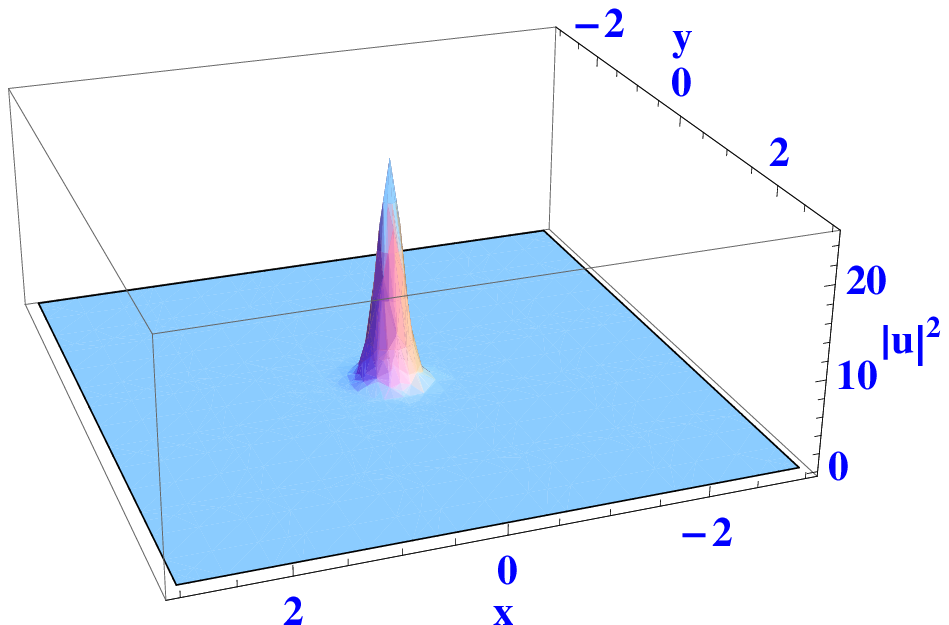}}
\subfigure[]{\label{abc1}\includegraphics[height=3.7 cm, width=4.5 cm]{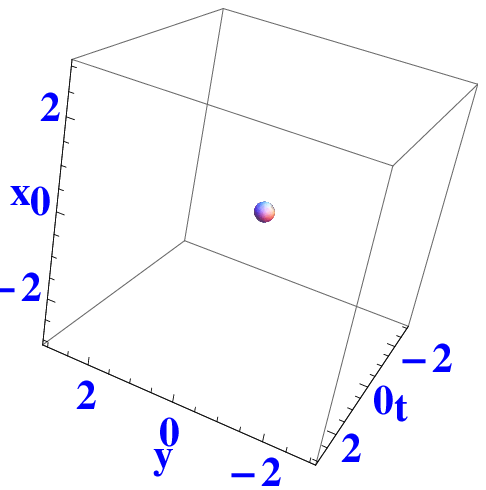}}
\end{center}
\caption{(Color online) Profile of the tightly focussed beam formed during the evolution of HGBs. (a) Spatial profile (t=0). (b) Spatio-temporal profile. }
  \label{exm}
\end{figure*}
Even though the evolution of the beam is periodic where it gets self-trapped at equal intervals, the beam spreads after some periods. The evolution of spatial profile of the beam after first and second  periods of self-trapping for $\kappa$=0.5 and $\kappa$=2.0 are depicted in Figs. (3(k)- 3(l)) and Figs. (6(k)- 6(l)), respectively. The beam gets significantly spread out in successive periods of self-trapping. When compared to the case of $\kappa$=0.5, the distortion of the beam is much high for $\kappa$=2.0.  This is attributed to the fact that  for large propagation distances, the impact of cubic nonlinearity is not strong enough to compete with dispersive effects. Influence of higher order \cite{hig} or saturable \cite{sat1, sat2} nonlinearity may lead to more stable self-trapped propagation of the beam.
\section{Influence of $\kappa$ on critical power}
\begin{figure*}
\begin{center}
\includegraphics*[height=4.5cm, width = 6cm]{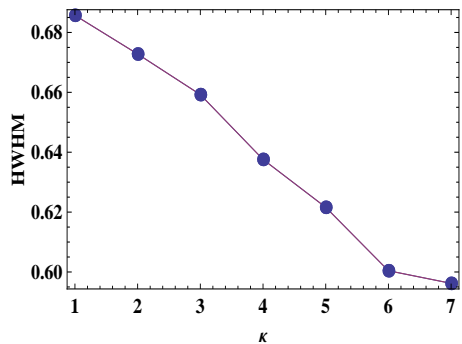}
\caption {(Color online) Half width at half maxima of the central bright spot versus $\kappa$.}
\label{new2}
\end{center}
\end{figure*}
The propagation properties of HGB is highly influenced by the input wave power, the soliton-like propagation and self trapping occur at a particular input wave power. {\color {blue}This important parameter is called the critical power ($P_c$) and it is a function of  $\kappa$. Its dependance is portrayed by adopting both the variational analysis and numerical method in Fig. \ref{pcr}.} It is clear from the figure that ($P_c$) increases linearly with $\kappa$. At higher values of  $P_c$ the self focussing nonlinearity is strong enough to compete with diffractive and dispersive effects, thereby the self-trapping of the beam will occur comparatively at low periods. Hence at higher values of $\kappa$ the self trapping period of the beam is low. The evolutions of HGB such as  the disappearance of single ringed intensity pattern and the appearance of intensity patterns with central maximum bright spot are achievable with less propagation distance in metamaterials for higher values of $\kappa$, which can be noted from Fig. \ref{abc1112b}. On the other hand, metamaterials with low value of $\kappa$  can preserve single ring intensity distribution over long propagation distances without focusing (Fig. \ref{m1sp5}).
\section{Tight focusing}
\begin{figure*}
\begin{center}
\includegraphics*[height=4.5cm, width = 6cm]{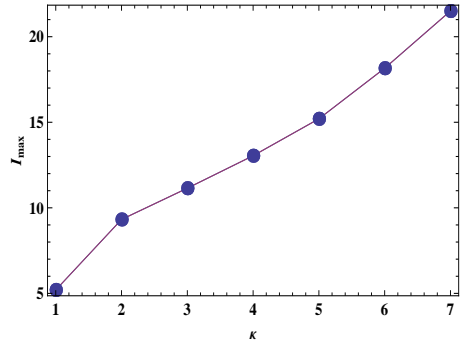}
\caption {(Color online) Maximum intensity of the central bright sport versus $\kappa$.}
\label{new1}
\end{center}
\end{figure*}
{\color {blue}It is quite clear from the study on periodic intensity evolution of the HGB that at the middle of each period the beam evolves to highly focussed structure with a central bright spot. An example for such a beam profile is illustrated in Fig. \ref{exm}.} {\color {blue}As the value of $\kappa$ increases the focussing of HGB is significantly sharp with narrow peak and high intensity. This is due to the consequence of increase in the critical power of self-trapping as the value of $\kappa$ increases. In Fig. \ref{new2} we have depicted the change of half width at half maxima of the central bright spot with respect to $\kappa$ numerically. It is clear from the figure that as the value of $\kappa$ increases the half width at the half maximum (HWHM) of the central bright spot decreases. Interestingly the numerical plot between peak power of the bright spot and $\kappa$ as depicted in Fig. \ref{new1} shows that the peak power of the bright spot increases with $\kappa$.}
\par
This study revives the possibility of generating highly focussed beam from HGB using metamaterials.  It is interesting
to note that metamaterials are engineered entities and one can tune their parameters at will. This freedom of metmaterials make this theoretical study quite meaningful in terms of generating highly focussed beams by controlling the dimensionless dispersion coefficient. The highly focussed beams may find interesting applications in terms of  trapping and manipulating nanosized dielectric spheres with high refractive index in the focal region.
\section{Conclusion}
In this paper, we have investigated the propagation properties of HGBs in metamaterials using the variational method. It is found that the self-trapped propagation of HGBs and the evolution of the beam are highly  influenced by the dimensionless dispersion coefficient.  The transformations of HGB such as  the disappearance of single ringed intensity pattern and the appearance intensity patterns with the central maximum with central bright spot are achievable over less propagation distance in metamaterials with higher values of $\kappa$. On the other hand, metamaterials with low value of $\kappa$  can preserve single ring intensity distribution over long propagation distances without focusing. When the strength of dispersion over diffraction increases the HGBs evolve to tightly focussed beam with high peak intensity at the center. This phenomenon of tight focussing is found to have interesting application in the trapping of nanosized particles with high index of refraction in the focal region. We believe that our theoretical results will help to stimulate new experiments with metamaterials.
\section{Acknowledgement}
 The work of A.K.S. is supported by the University Grants Commission (UGC), Government of India, through a D. S. Kothari Post Doctoral Fellowship in Sciences. M.L. is supported by DST-SERB through a Distinguished Fellowship (Grant No. SB/DF/04/2017).


\begin{thebibliography}{30}
\bibitem{tak}T. Kuga, Y. Torii, N. Shiokawa, and T. Hirano, Phys. Rev. Lett. 78 (1997) 4713–4716.
\bibitem{Jia}J. Yin, Y. Zhu, W. Jhe, and Z. Wang, Phys. Rev. A. 58 (1998) 509.
\bibitem{sinh}Y. Cai, X. Lu, and Q. Lin, Opt. Lett. 28 (2003) 1084-1086.
\bibitem{Uni}D. Deng, H. Yu, S. Xu, J. Shao and Z. Fan, Opt. Commun. 281 (2008) 202-209.
\bibitem{Ms}M.S. Sodha, S.K. Mishra and S. Misra, Laser and Particle Beams 27 (2009) 57–68.
\bibitem{Ms1}I. Gerdova, X. Zhang, and A. Hache, J. Opt. Soc.  Am. B 23 (2006) 1934-1937 .
\bibitem{nonlocal}Y. Z. Jun, Lu Da Quan, Hu Wei, Zheng Yi Zhou and Gao Xing Hui, Chin. Phys. B 19 (2010) 124212.
\bibitem{nonlocal1}Z. Dai, Z. Yang, S. Zhang, Z. Pang and K. You, Laser Phys. 25 (2015) 025401.
\bibitem{G1}Z. Liu, H. Zhao, J. Liu, Jie Lin, M. Ashfaq Ahmad, and S. Liu, Opt. Lett. 32 (2007) 2076-2078.
\bibitem{G2}Z. Liu, J. Dai, X. Sun, and S. Liu, Opt. Exp. 16 (2008) 19926-19933.
\bibitem{G3}N. A. Chaitanya, M. V. Jabir, J. Banerji and G. K. Samanta, Scientific Reports 6 (2016) 32464.
\bibitem{G4}Y. Nie, X. Li, Junli Qi, H. Ma, J. Liao, J. Yang and W. Hua, Opt.  Laser Technol. 44 (2012) 384-389.
 \bibitem{SH}G. Purohit, P. Rawat, and R. Gauniyal, Phys.  Plasmas 23 (2016) 013103.
 \bibitem{trap}C. Liang, Z. Gang Wang and X. Hui Lu, Phys. Lett. A 363 (2007)  502.
  \bibitem{trap1}B. Tang, Y. Li, X. Zhou, L. Huang and X. Lang, Optik 127 (2016) 6446-6451.
\bibitem{trap2}Z. Liu , X. Wang and K. Hang, Scientific Reports 9 (2019) 10187.
\bibitem{mt1}V. M. Agranovich, Y. R. Shen, R. H. Baughman, and A. A. Zakhidov, Phys. Rev. B 69 (2004) 165112.
\bibitem{mt2}N. Papasimakis, V. A. Fedotov, N. I. Zheludev, and S. L. Prosvirnin, Phys. Rev. Lett. 101 (2008) 253903.
\bibitem{mt3}A. Poddubny, I. Iorsh, P. Belov and Y. Kivshar, Nature Photonics 7 (2013) 948–957.
\bibitem{Ramakrishna}S. A. Ramakrishna, Rep. Prog. Phys. 68 (2005) 449521.
\bibitem{Caloz}C. Caloz and T. Itoh, "Electromagnetic Metamaterials: Transmission Line Theory and Microwave Applications" (Wiley Interscience, 2006).
\bibitem{opt}Vladimir M. Shalaev, Nature Photonics 1 (2007) 41–48.
{\color {blue}\bibitem{Pendry1} J. B. Pendry,  A. J. Holden, W. J. Stewart, and I. Youngs, Phys. Rev. Lett. 76 (1996) 4773.
\bibitem{Pendry2}J. B. Pendry, A. J. Holden, D. J. Robbins, and W. J. Stewart, J. Phys.: Condens. Matter. 10  (1998) 4785.
\bibitem{Pendry3}J. B. Pendry, A. J. Holden, D. J. Robbins, and W. J. Stewart, IEEE Trans. Microw. Theory Tech. 47 (1999), 2075.
\bibitem{Smith} D. R. Smith, W. J. Padilla, D. C. Vier, S. C. Nemat-Nasser, and S. Schultz, Phys. Rev. Lett. 84 (2000) 4184 .
\bibitem{roppo}V. Roppo, C. Ciraci, C. Cojocaru and M. Scalora, J. Opt. Soc. Am. B 27  (2010) 1671-1679.
\bibitem{Ilya} I. V. Shadrivov, S. K. Morrison, and Y. S. Kivshar, Opt. Exp. 14  (2006) 9344.
\bibitem{Zharov}A. A. Zharov, I. V. Shadrivov, and Y. S. Kivshar, Phys. Rev. Lett. 91  (2003) 037401.}
\bibitem{n1}A. Chowdhury and John A. Tataronis, Phys. Rev. E 75 (2007) 016603.
\bibitem{n2}H. Suchowski, K. OBrien, Zi Jing Wong, A. Salandrino, X. Yin and X. Zhang, Science 342 (2013) 1223-1226.
\bibitem{n3}S. Lan, L. Kang, D. T. Schoen, S. P. Rodrigues, Y. Cui, M. L. Brongersma and W. Cai, Nature Materials 14 (2015) 807–811.
\bibitem{n11}Ilya V. Shadrivov, Alexander A. Zharov, and Yuri S. Kivshar, J. Opt. Soc. Am. B 23 (2006) 529-534.
\bibitem{n12} A. Ciattoni, C. Rizza, and E. Palange, Opt. Lett. 35 (2010) 2130-2132.
\bibitem{n13}Y. Xiang, X. Dai, S. Wen, Jun Guo, and D. Fan, Phys. Rev. A 84 (2011) 033815.
\bibitem{n14}P. Yen Chen, M. Farhat, and A. Alu, Phys. Rev. Lett. 106 (2011) 105503.
\bibitem{n15}S. Wen, Y. Wang, W. Su, Y. Xiang, X. Fu, and D. Fan, Phys. Rev. E  73 (2006) 036617.
\bibitem{wenf}S. Wen, Y. Xiang, X. Dai, Z. Tang, W. Su, and D. Fan, Phys. Rev. A 75 (2007) 033815.
\bibitem{wen1}J. Zhang, S. Wen, Y. Xiang, Y. Wang, and H. Luo, Phys. Rev. A 81 (2010) 023829.
{\color {blue}\bibitem{CJ}C. J. Tang, P. Zhan, Z. S. Cao, J. Pan, Z. Chen and Z. L. Wang, Phys. Rev. B. 83 (2011) 041402(R).}
\bibitem{beam}Y. Cai, X. Lu and Q. Lin, Opt. Lett. 28 (2003) 1084.
\bibitem{beam2}S. Konar, S. Jana and M. Mishra, Opt. Commun. 255 (2005) 114–129.
\bibitem{hig}K. Dimitrevski, E. Reimhult, E. Svensson, A.Ohgren, Dan Anderson, A. Berntson, M. Lisak, M. L. Quiroga-Teixeiro, Phys. Lett. A 248  (1998) 369.
\bibitem{sat1}M. Karlsson, Phys. Rev. A 46 (1992) 2726.
\bibitem{sat2}Z. Jovanoski and R. A. Semmut, Phys. Rev. E 50 (1994) 4087.
\end{thebibliography}
\end{document}